\documentclass[10pt,a4paper,notitlepage,prd,showpacs,preprintnumbers,superscriptaddress,nofootinbib]{revtex4-1}
\usepackage{float}
\usepackage{amsmath}
\usepackage{cancel}
\usepackage{graphicx}

\makeatletter

\usepackage{bm}
\usepackage{tensor}
\usepackage{comment}
\usepackage{ifpdf}
\usepackage{slashed}
\usepackage{color}
\usepackage[mathscr]{eucal}

\ifpdf
  \usepackage{graphicx}     
  \usepackage[bookmarksopen,colorlinks=true,linkcolor=bblue,citecolor=bblue,urlcolor=ppink]{hyperref}
\else     
  \usepackage[dvipdfmx]{}     
  \usepackage[dvipdfmx,bookmarksopen,colorlinks=true,linkcolor=bblue,citecolor=ppink,urlcolor=darkred]{hyperref}
\fi

\definecolor{red}{rgb}{1,0,0}
\definecolor{darkred}{rgb}{0.6,0,0}
\definecolor{darkgreen}{rgb}{0.992447,0.623778,0.034597}
\definecolor{ppink}{rgb}{1,0.4,0.4}
\definecolor{bblue}{rgb}{0.284602,0.317763,0.963947}

\makeatother

\begin{document}

\preprint{IPMU19-0187}
\title{Revisiting oscillon formation in KKLT scenario}

\author{Shinta Kasuya}
\affiliation{Department of Mathematics and Physics, Kanagawa University, Kanagawa 259-1293, Japan}

\author{Masahiro Kawasaki}
\affiliation{ICRR, University of Tokyo, Kashiwa, 277-8582, Japan}
\affiliation{Kavli IPMU (WPI), UTIAS, University of Tokyo, Kashiwa, 277-8583,
Japan}

\author{Francis Otani}
\affiliation{ICRR, University of Tokyo, Kashiwa, 277-8582, Japan}
\affiliation{Kavli IPMU (WPI), UTIAS, University of Tokyo, Kashiwa, 277-8583,
Japan}

\author{Eisuke Sonomoto}
\affiliation{ICRR, University of Tokyo, Kashiwa, 277-8582, Japan}
\affiliation{Kavli IPMU (WPI), UTIAS, University of Tokyo, Kashiwa, 277-8583,
Japan}

\begin{abstract}
\noindent 
KKLT scenario has succeeded in stabilizing the volume modulus and constructing metastable 
de Sitter (dS) vacua in type IIB string theory. We revisit to investigate the possibility 
of the oscillon (or I-ball) formation in the KKLT scenario when the volume modulus is initially 
displaced from the dS minimum. Special attention is paid to physically realistic initial 
conditions of the volume modulus, which was not taken in the literature. Using lattice 
simulations, we find that oscillons do not form unless the volume modulus is initially 
placed at very near the local maximum, which requires severe fine-tuning.
\end{abstract}

\maketitle



\section{Introduction}
\label{sec:introduction} 


String theory, in general, contains numerous scalar fields, including the so-called moduli fields 
which determine the shape and size of extra dimensions. It was difficult for some time to 
achieve realistic phenomenological consequences from them since stabilization of the 
compactification of extra dimensions (i.e., moduli) remained to be an unsolved poser. 
Among them, it was most difficult to stabilize the volume modulus, also called the K\"{a}hler 
modulus. However, a possible solution to this problem was first proposed in \cite{Kachru:2003aw}, 
known as the KKLT scenario, eventually followed by the solution such as the Large Volume Scenario 
\cite{Balasubramanian:2005zx}. These scenarios succeeded not only in stabilizing 
the volume moduli in type IIB string theory, but also in enabling to construct metastable 
de Sitter (dS) vacua, opening the doors to explain observational cosmology by string theory.

It is believed that these moduli might change the course of the history of the standard cosmology 
in the early Universe by adding an extra matter (i.e., moduli) dominated era, which can lead to 
additional contributions of dark radiation \cite{Cicoli:2012aq, Higaki:2012ar}, baryogenesis \cite{Allahverdi:2016yws} and non-thermal productions of dark matter  \cite{Allahverdi:2013noa}. 
It is also natural to ask what the phenomenological consequences are when the volume moduli are
displaced from the dS minimum in these scenarios. Thus, we focus on their cosmological dynamics in 
the KKLT scenario.

The phenomenon that we are particularly interested in is the production of oscillons
\cite{Bogolyubsky:1976yu, Gleiser:1993pt, Copeland:1995fq}. 
Oscillons are spatially localized and long-lived non-topological solitons that could form when 
the scalar field $\phi$ oscillates around the minimum of a certain potential.
For them to form, the potential needs to be slightly shallower than quadratic, at least 
in some regions of the field amplitude, and the perturbations of the scalar field must 
grow sufficiently as well. Oscillons are known to appear in various types of potentials, 
such as inflaton potentials \cite{Amin:2011hj,McDonald:2001iv, Lozanov:2017hjm, Hasegawa:2017iay},
axion-like potentials \cite{Kolb:1993hw, Hong:2017ooe, Kawasaki:2019czd}, etc.
It is also noteworthy that the stability of the oscillon is ensured by an adiabatic invariant 
$I$~\cite{Kasuya:2002zs, Kawasaki:2015vga}, since it can be defined as the scalar configuration
of the energy minimum with fixed $I$, which is why they are also called I-balls.
When oscillons are generated, they can have a large impact on the cosmological evolution of the Universe:
they could dominate the energy density of the universe and delay thermalization 
\cite{Gleiser:2006te}. In particular, it could be a source of 
characteristic gravitational waves \cite{Zhou:2013tsa,Antusch:2017vga}, which may give some 
indications for string theories.
 
The oscillon formation in models based on string theory including the KKLT was examined before in \cite{Antusch:2017flz} to some extent. However, we find that the initial conditions of the 
volume modulus is not natural and appropriate in that paper. We therefore consider physically 
sensible initial values of the volume modulus and the Hubble parameter $H$, which would change 
the results of the simulations, and lead to different conclusions.
 
The paper is organized as follows. In Sec.~\ref{sec:KKLT -model}, we briefly review the potential
of the volume modulus in the KKLT scenario. In Sec.~\ref{sec:Floquet Analysis}, we reexamine the 
growth of the instability using Floquet analysis following the procedures in \cite{Antusch:2017flz}. 
In Sec.~\ref{sec:Lattice Simulation}, we present the results of lattice simulations. 
The last section is devoted to conclusions.
Throughout the paper we set the reduced Planck mass to $M_{P} = 1$.


\section{KKLT model}
\label{sec:KKLT -model}


In this section, we derive the potential of the volume modulus $T$ in the KKLT scenario, which is used
in the lattice simulations of the oscillon formation. 
The relevant K\"ahler potential and superpotential in the simplest case are
\begin{equation}
    K=-3\ln(T+\bar{T}), \quad W=W_{0}+Ae^{-\kappa T},
    \label{Kahlerpot}
\end{equation}
respectively, where $W_0$ is the superpotential after fluxes stabilize the axion–dilaton and 
complex structure moduli. The second term in the superpotential comes from gaugino condensation 
of $N$ D7-baranes with $\kappa=2\pi/N$, or wrapped Euclidean D3 brane instantons with $\kappa=2\pi$.
In either case, $A$ is independent of the volume modulus and $\kappa\sim{\cal O}(1)$.
Although the volume modulus is generically a complex scalar field ($T=\sigma+i\theta$), here we 
simply set the imaginary component $\theta$ to zero following \cite{Kachru:2003aw}. 
Then the potential can be obtained as
\begin{equation}
V(\sigma) = \frac{\kappa Ae^{-\kappa\sigma}}{2\sigma^{2}}
\left[ (1+\frac{\kappa}{3}\sigma)Ae^{-\kappa\sigma}+W_{0} \right].
\end{equation}

This potential has the AdS vacuum, and must be lifted to a metastable dS vacuum by 
adding a small positive uplifting term as
\begin{equation}
    \delta V=\frac{D}{\sigma^{3}},
\end{equation}
which is achieved, for example, by the effects from anti-D3 branes where $D$ is a positive constant. 
The value of $D$ is fine-tuned so that the potential of 
the minimum is slightly positive, realizing the current stage of acceleration of the Universe. 
Another example of uplifting terms could arise using D7 branes \cite{Burgess:2003ic}. 
The exponent of $\sigma$ of the uplifting term depends on each 
mechanism. The total Lagrangian including the kinetic term is finally given by
\begin{equation}
    {\cal L} = 
    \sqrt{-g}\left[ \frac{3}{4\sigma^2}(\partial \sigma)^2
    -\left(V(\sigma) +\frac{D}{\sigma^{3}}\right) \right],
\end{equation}
in the FLRW background
\begin{equation}
    ds^{2}=dt^{2}-a^{2}(t)d{\bf x}^{2},
\end{equation}
with $a(t)$ being the scale factor, and $g$ is the determinant of the metric $g_{\mu\nu}$.

Introducing a canonically normalized scalar field $\phi$ as\footnote{%
We correct the factor which differs by $\sqrt{2}$ from that applied in \cite{Antusch:2017flz}.
Hence, the following potential is slightly different from theirs.
}
\begin{equation}
    \phi=\frac{\sqrt{6}}{2}\ln(T+\bar{T})=\frac{\sqrt{6}}{2}\ln(2\sigma),
\end{equation}
we obtain the Lagrangian as
\begin{equation}
    {\cal L}=\sqrt{-g}\left[\frac{1}{2}(\partial \phi)^{2}-V(\phi)\right],
\end{equation}
where
\begin{equation}
V(\phi)=\frac{\kappa A\exp\left(-\kappa\frac{e^{\sqrt{6}\phi/3}}{2}\right)}{2(\frac{e^{\sqrt{6}\phi/3}}{2})^2}
\left\{ \left[1+\frac{\kappa}{3}\left(\frac{e^{\sqrt{6}\phi/3}}{2}\right)\right]
A \exp\left(-\kappa\frac{e^{\sqrt{6}\phi/3}}{2}
\right)+W_{0} \right\}+\frac{D}{\left(\frac{e^{\sqrt{6}\phi/3}}{2}
\right)^{3}}.
\label{KKLTpt}
\end{equation}

\begin{figure}[h]
  \begin{center}
    \includegraphics[width=12.0cm]{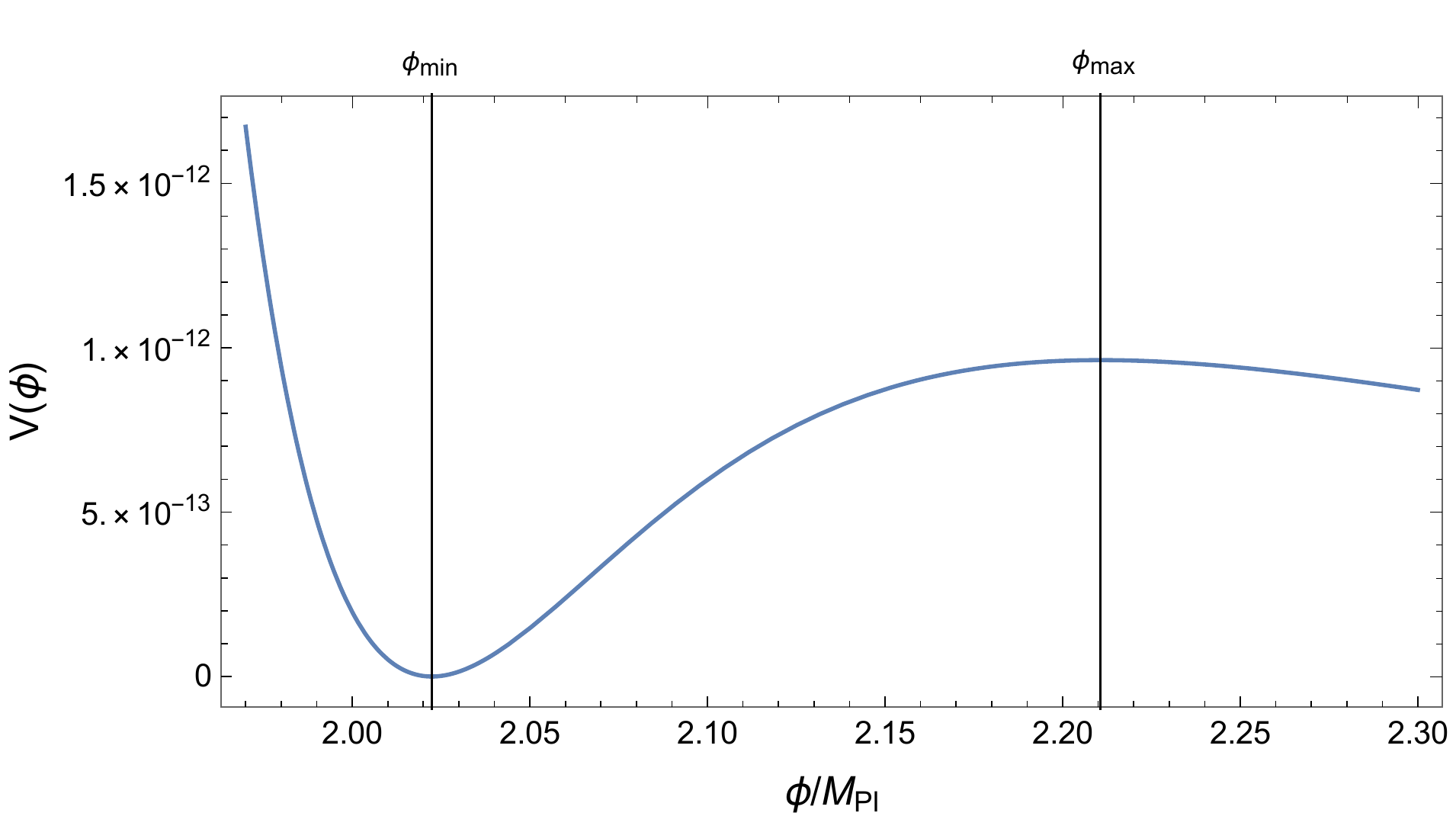}
    \caption{Potential of KKLT described by equation \eqref{KKLTpt} in the case of 
    $W_{0}=-10^{-5}, A=10, \kappa=2\pi$ and $D=3.17\times10^{-11}$. $\phi_{\text{min}}$ and 
    $\phi_{\text{max}}$ are the field values at the minimum and the local maximum of 
    the potential, respectively.}
    \label{KKLTpotential}
  \end{center}
\end{figure}

In our numerical simulations, we adopt the following parameter ranges,
\begin{equation}
    -10^{-5} \leq W_{0} \leq -10^{-12}, \quad \quad
    1 \leq A \leq 10, \quad \quad
    1 \leq \kappa \leq 2\pi,
    \label{parameterrange}
\end{equation}
which are the same as used in \cite{Antusch:2017flz} by the same reasons explained there. 
Figure \ref{KKLTpotential} shows an example of the KKLT potential.


\section{Floquet Analysis}
\label{sec:Floquet Analysis}


We first conduct the Floquet analysis in this section to discuss the growth of amplitudes of the 
perturbations, following \cite{Antusch:2017flz}. This analysis can be applied when the 
fluctuations of the scalar field are small compared to the homogeneous and oscillating background 
field. Let us write the scalar field $\phi$ with the homogeneous background and the perturbation as
\begin{equation}
	\phi(t,{\bf x})=\phi(t)+\delta\phi(t,{\bf x}).
\end{equation}
The equation of motion of the homogeneous component is
\begin{equation}
	\ddot{\phi}(t)+3H\dot{\phi}(t)+V'(\phi(t))=0.
\label{homogeneouseq}
\end{equation}
For the perturbation, we decompose it into Fourier modes $\delta\phi_{k}$:
\begin{equation}
\delta\phi(t,{\bf x})=\int\frac{d^{3}k}{(2\pi)^{3}}\delta\phi_{k}(t)e^{-i{\bf k}{\bf x}}.
\end{equation}
The Fourier modes then obey the following equation:
\begin{equation}
\delta\ddot{\phi}_{k}+3H\delta\dot{\phi}_{k}+\left(\frac{k^{2}}{a^{2}(t)}
+V''(\phi(t))\right)\delta\phi_{k}=0.\label{perturbationeq}
\end{equation}
In the Floquet analysis, we ignore the expansion of the Universe and use the Minkowski space. 
Hence, equations \eqref{homogeneouseq} and  \eqref{perturbationeq} reduce respectively to
\begin{align}
\ddot{\phi}(t)+V'(\phi(t))&=0, \label{homoeq}\\
\delta\ddot{\phi}_{k}+\left[k^{2}+V''(\phi(t))\right]\delta\phi_{k}&=0. \label{perturbeq}
\end{align}
According to the Floquet theorem, the solution of equation \eqref{perturbeq} is given by
\begin{equation}
\delta\phi_{k}(t)=P_{+}(t)e^{\mu_{k}t}+P_{-}(t)e^{-\mu_{k}t},
\end{equation}
where $\mu_{k}$ is the Floquet exponent and $P_{\pm}$ are periodic functions with the same
period as the oscillation period of the background field, which we denote by $T$. 
Therefore, Re$(\mu_{k})\neq 0$ indicates instability growth of the fluctuation modes known 
as parametric resonance\cite{Shtanov:1994ce, Kofman:1995fi, Kofman:1997yn}.
The Floquet exponents can be obtained as follows. We solve the two 
equations \eqref{homoeq} and \eqref{perturbeq} simultaneously from $t = 0$ to $t = T$ for 
two sets of orthogonal initial conditions:
\begin{equation}
	\delta\phi_{k,1}(0)=1, \ \delta\dot{\phi}_{k,1}(0)=0 \quad {\rm and} \quad  
	\delta\phi_{k,2}(0)=0, \ \delta\dot{\phi}_{k,2}(0)=1.
\end{equation}
Floquet exponents are then computed by
\begin{equation}
{\rm Re}(\mu_{k}^{\pm})=\frac{1}{T}\ln\left| \frac{1}{2} \left( \delta\phi_{k,1}+ \delta\dot{\phi}_{k,2}
\pm\sqrt{\left( \delta\phi_{k,1}-\delta\dot{\phi}_{k,2}\right)^2
+4\delta\phi_{k,2}\delta\dot{\phi}_{k,1}} \right) \right|,
\end{equation}
where all the quantities within the logarithm are evaluated at $t=T$ (see \cite{Amin:2014eta} 
for more details). We compare the real part of the Floquet exponent to the mass of the modulus at 
the potential minimum,
\begin{equation}
    m\equiv \left.\sqrt{\frac{\partial^{2}V(\phi)}{\partial \phi^{2}}}\right|_{\rm min},
\end{equation}
since the timescale of the oscillation can be considered as $m^{-1}$ for most of the range of the
initial field amplitude $\phi_{\rm initial}$.

\begin{figure}[t]
  \begin{center}
    \begin{tabular}{c}
      \begin{minipage}{0.45\hsize}
        \begin{center}
          \includegraphics[width=\textwidth]{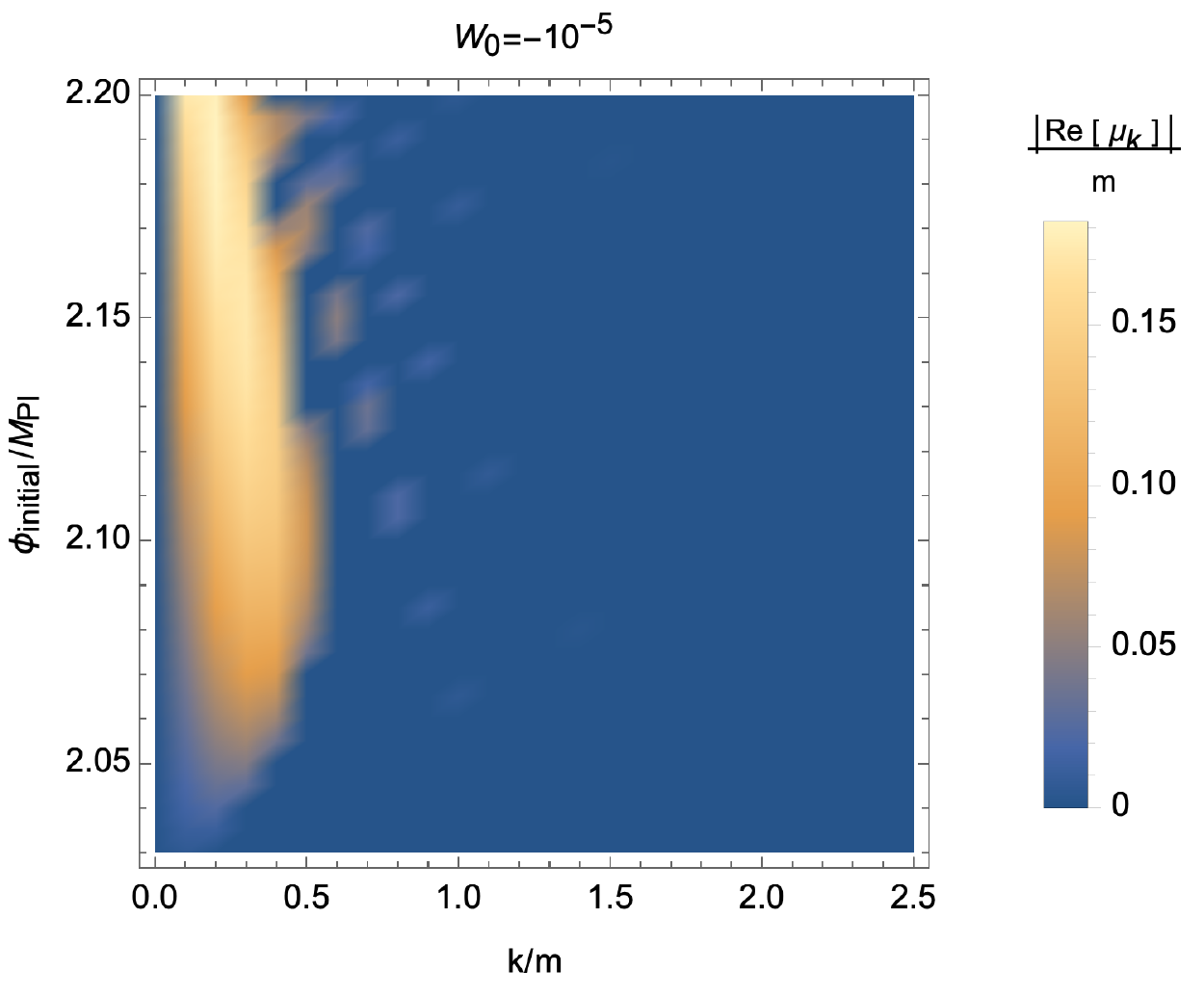}       
        \end{center}
      \end{minipage}
      \begin{minipage}{0.45\hsize}
        \begin{center}
          \includegraphics[width=\textwidth]{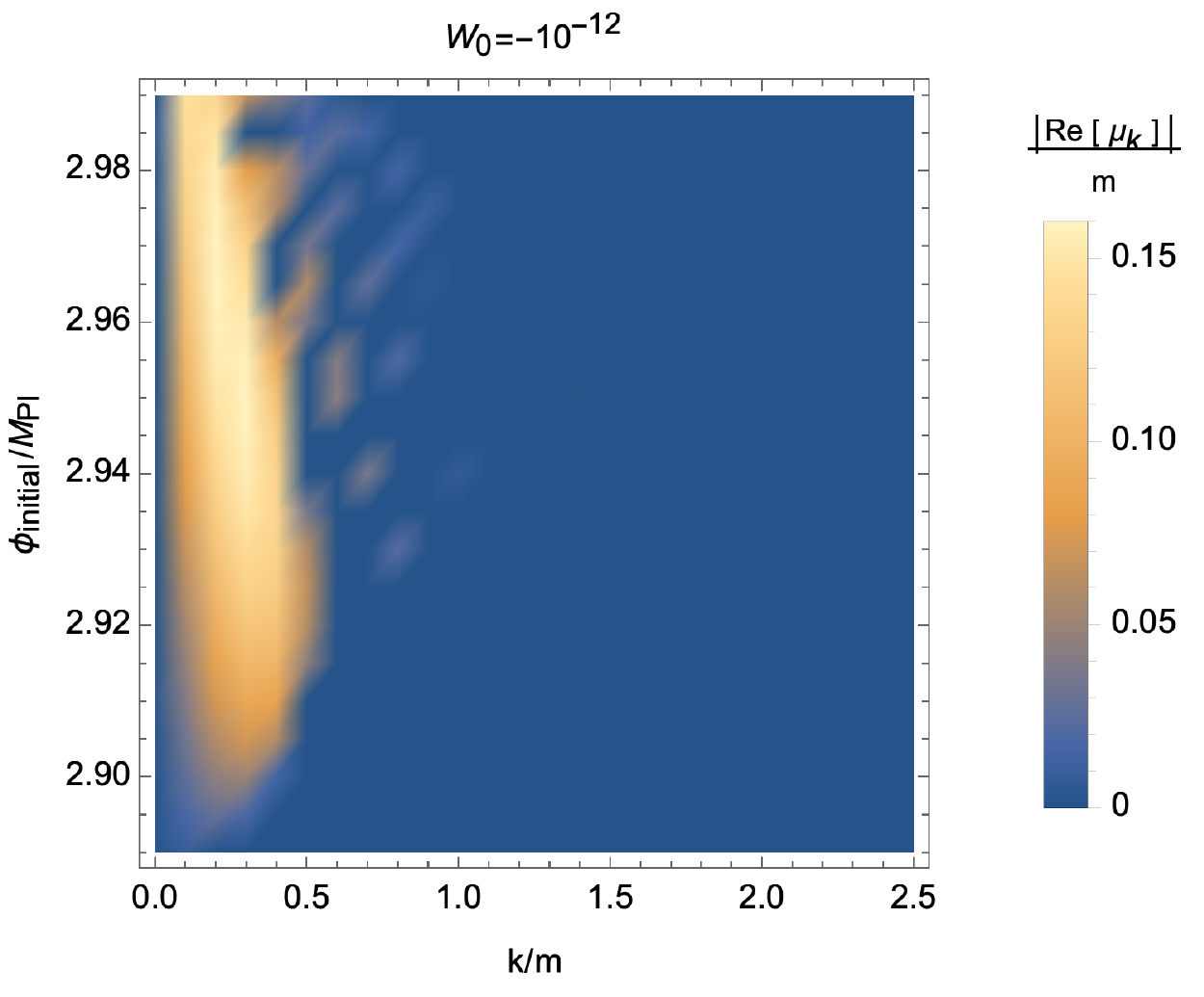}
        \end{center}
      \end{minipage}
    \end{tabular}
    \caption{Floquet exponents of the KKLT model in units of the modulus mass 
    for $W_{0}=-10^{-5}, A=10, \kappa=2\pi$ (left) and $W_{0}=-10^{-12}, A=10, \kappa=2\pi$ (right). }
    \label{FA}
  \end{center}
\end{figure}

Figure \ref{FA} shows the results of the Floquet analysis for $W_{0}=-10^{-5}, A=10, \kappa=2\pi$ (left) 
and $W_{0}=-10^{-12}, A=10, \kappa=2\pi$ (right).  In both cases, there exists a broad instability band 
at $k\lesssim 0.5m$, but the value of the exponent there is ${\rm Re}(\mu_{k})/m\sim0.1$, which is 
generically not enough for the fluctuation modes to grow. Thus, it seems hard for the oscillon formation to take place.

However, the tachyonic instability could occur when the modulus field is initially placed near 
the local maximum of the potential, where the curvature of the potential (the second derivative of the potential $V''(\phi)$) is negative. Then the modes 
with wavenumbers with $k^{2}+V''(\phi)<0$ grow exponentially, which might result in the non-linear 
interactions among the fields leading to the formation of oscillons.
We will see if this could happen in the lattice simulations in the next section.
  

\section{Lattice Simulation}
\label{sec:Lattice Simulation}

\subsection{Setups}

We study the non-linear dynamics of the scalar field of the KKLT model by using a modified 
version of Lattice-Easy~\cite{Felder:2000hq}, solving the following equation on the lattices:
\begin{equation}
    \ddot{\phi}+3\frac{\dot{a}}{a}\dot{\phi}-\frac{1}{a^{2}}\nabla^{2}\phi
    +\frac{\partial V(\phi)}{\partial \phi}=0.
    \label{eqmotionlattice}
\end{equation}

The volume modulus starts to roll down the potential 
when the Hubble parameter becomes smaller than the curvature of the potential. 
We can thus naturally set the initial value of the homogeneous field, $\phi_{\rm init}$, 
and the Hubble parameter at that time, by
\begin{equation}
H_{\rm initial}=m_{\rm eff}(\phi_{\rm initial}),
\label{H_{init}}
\end{equation}
where we define as
\begin{equation}
    m_{\text{eff}}(\phi)\equiv \sqrt{\left|\frac{\partial^{2}V(\phi)}{\partial \phi^{2}}\right|},
    \label{meff}
\end{equation}
together with the initial value of the time derivative of the field being $\dot{\phi}_{\text{initial}}=0$.
Notice that this is the crucial difference from those adopted in \cite{Antusch:2017flz}, 
in which they set 
\begin{equation}
H_{\rm initial}=\sqrt{\frac{V(\phi_{\rm initial})}{3}} \quad {\rm for} \quad
\phi_{\rm initial}=\phi_{\rm min}+\frac{\phi_{\rm max}-\phi_{\rm inf}}{2}, \quad {\rm and} \quad
\dot{\phi}_{\text{initial}}=0,
\end{equation}
where $\phi_{\rm min}$, $\phi_{\rm max}$, and $\phi_{\rm inf}$ are the field values at
the minimun, maximun, and the inflection point of the potential, respectively.
Since $m_{\text{eff}}(\phi_{\rm intial})$ is larger than $\sqrt{V(\phi_{\text{initial}})/3}$ 
by about one order of magnitude for the majority of the field amplitudes, the field should have 
already rolled down if the latter had been adopted as the Hubble parameter. Therefore, the initial 
condition $\dot{\phi}_{\text{initial}}=0$ in \cite{Antusch:2017flz} is impossible to hold. 
As shown below, the naturally realized value of the Hubble parameter (\ref{H_{init}}) forbids the 
generation of oscillons for an even wider range of $\phi_{\text{initial}}$.

As for the initial fluctuations of the scalar field, quantum vacuum fluctuations are used whose 
magnitude and phase follow the Rayleigh distribution and the random uniform distribution, respectively 
(see \cite{Felder:2000hq,Polarski:1995jg,Khlebnikov:1996mc}).
We assume the Universe is matter-dominated and evolve the scale factor accordingly starting from $a=1$.

We perform simulations using the rescaled variables
\begin{equation}
    \phi_{pr}\equiv a\phi,  \quad \vec{x}_{pr}\equiv m\vec{x}
    , \quad dt_{pr}\equiv m\frac{dt}{a}
    \label{setting}
\end{equation}
on the two-dimensional lattices, since we need high enough resolution with reasonable computation time. 
The scale factor evolves in the program as
\begin{equation}
   a = \left(\frac{1}{2}m_{\rm eff}(\phi_{\rm initial})t_{pr} + 1 \right)^{2}.
\end{equation}
We set the simulation parameters as described in Table~\ref{Ta:params}, and adopt the model parameters
with the ranges shown in (\ref{parameterrange}) as mentioned.

\begin{table}[t]
    \centering
    \caption{
        Parameters adopted in our simulations.
    }
    \vspace{0.3cm}
    \begin{tabular}{ll}
        \hline \hline
        Box size & $L_{pr}=20$ \\ 
        Grid size & $N=1024^2$ \\ 
        Final time & $t_{f\_ pr}=60$ \\ 
        Time-step & $dt_{pr}=5\times 10^{-3}$ \\ 
        \hline \hline
    \end{tabular}
    \label{Ta:params}
\end{table}

\subsection{Results}

\begin{figure}[h]
  \begin{center}
    \begin{tabular}{c}
      \begin{minipage}{0.45\hsize}
        \begin{center}
          \includegraphics[width=\textwidth]{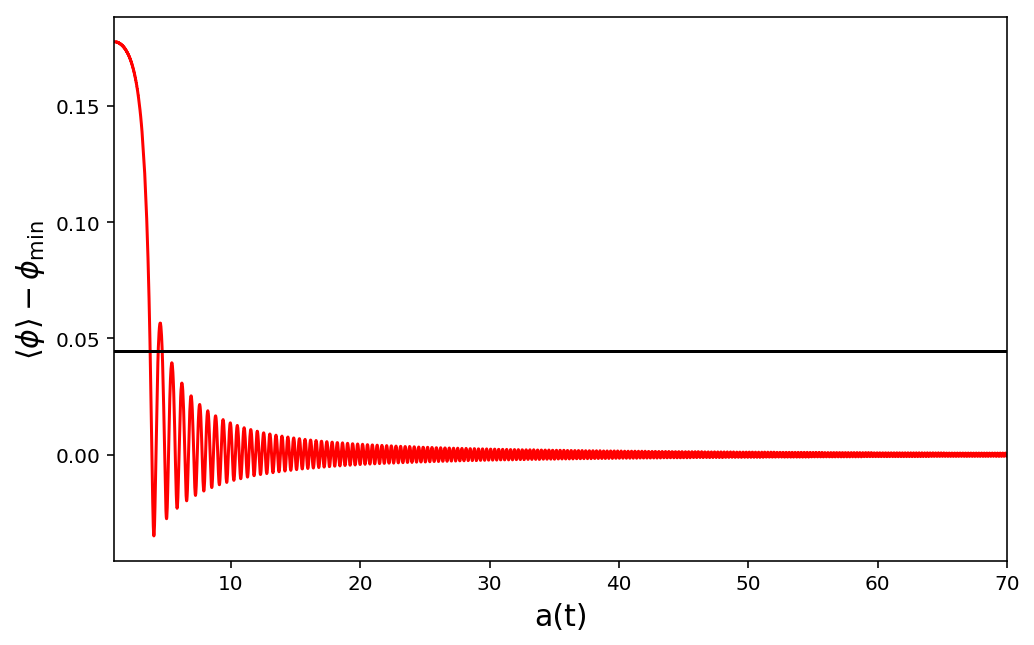}    
        \end{center}
      \end{minipage}
      \begin{minipage}{0.45\hsize}
        \begin{center}
          \includegraphics[width=\textwidth]{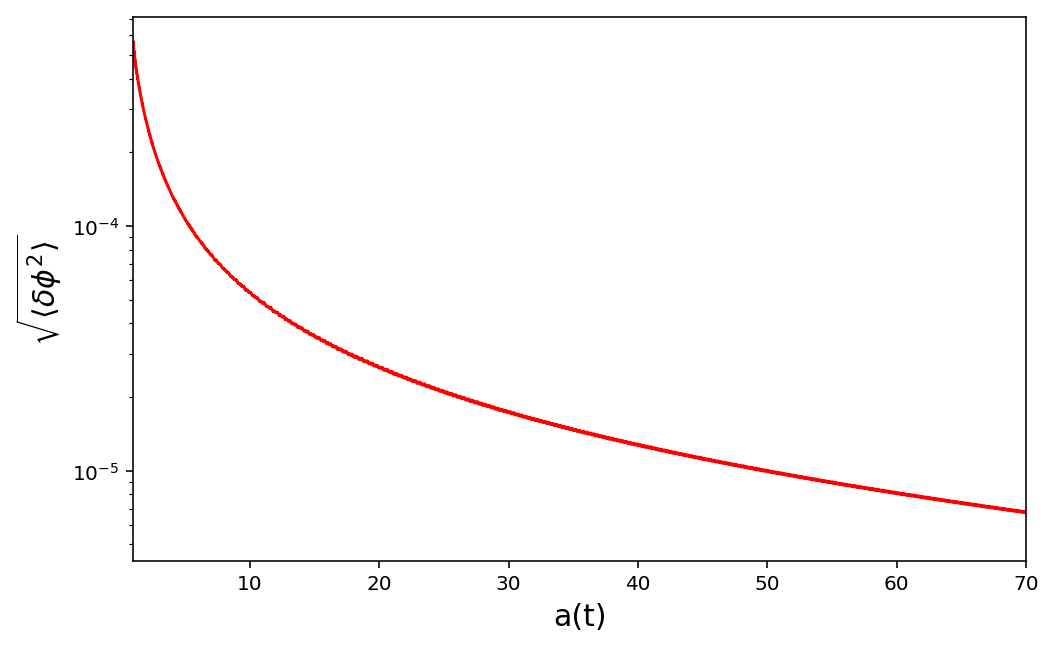}
        \end{center}
      \end{minipage}\\
      
        \begin{minipage}{0.45\hsize}
        \begin{center}
          \includegraphics[width=\textwidth]{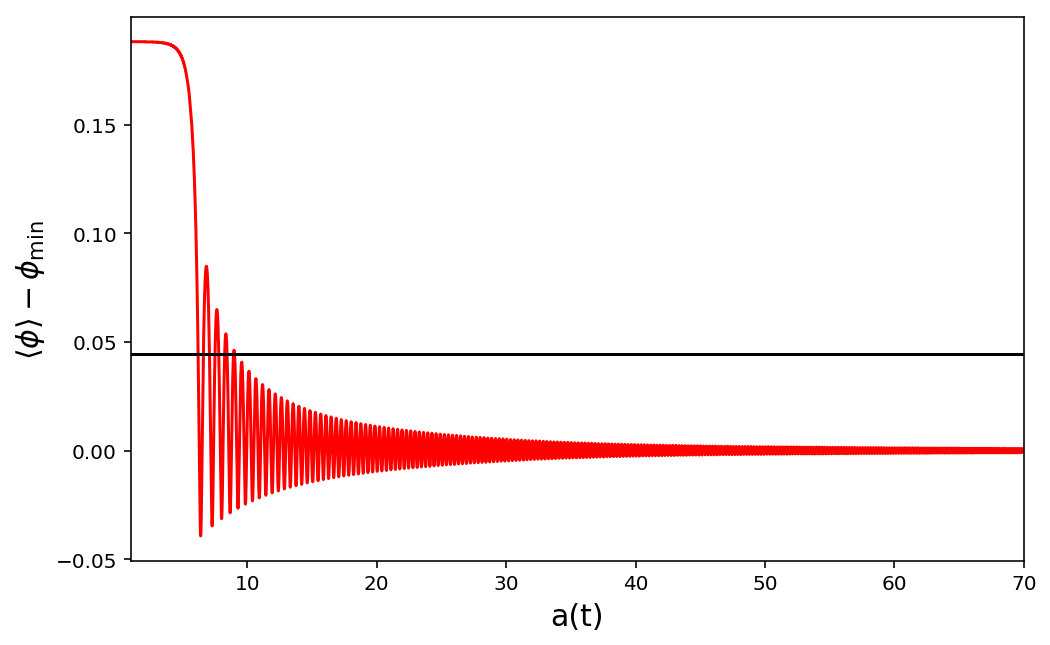}    
        \end{center}
      \end{minipage}
      \begin{minipage}{0.45\hsize}
        \begin{center}
          \includegraphics[width=\textwidth]{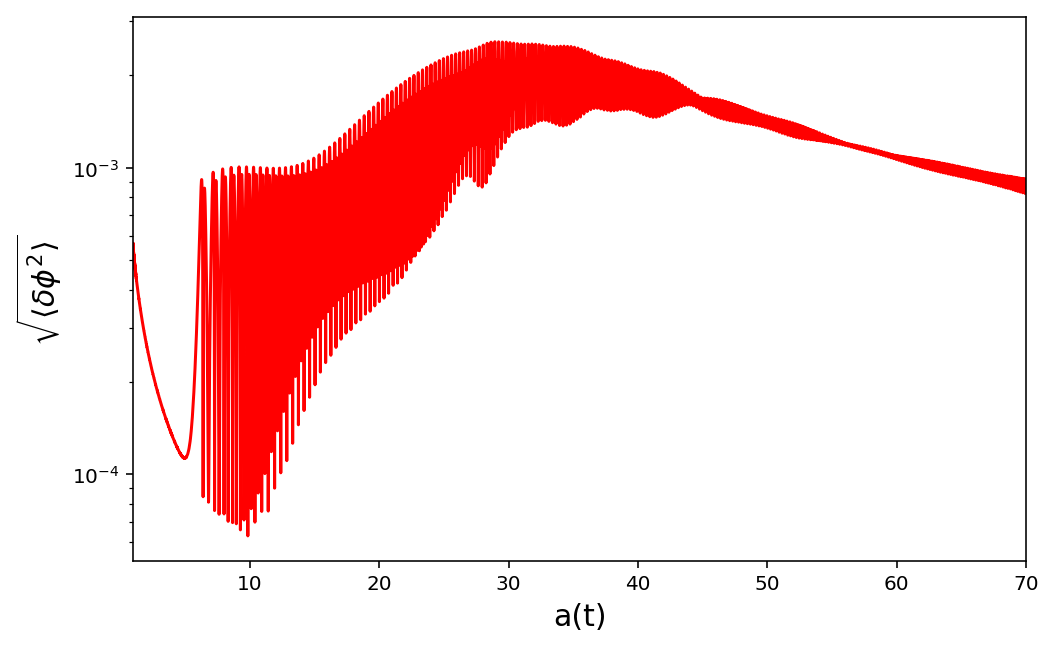}
        \end{center}
      \end{minipage}
    \end{tabular}
    \caption{
        Evolution of the mean $\langle \phi \rangle$ (left) and the variance 
        $\sqrt{ \langle\delta\phi^2 \rangle}$ (right) when $W_0=-10^{-5}, 
        A=10$, and $\kappa=2\pi$ as a function of the scale factor $a(t)$.
        The initial field values are taken to be $\phi_{\text{initial}}=2.2$ (top)
        and $\phi_{\text{initial}}=2.2105$ (bottom). 
        The solid black line indicates the amplitude of the field at 
        the inflection point of the potential.}
    \label{Fielddynamics}
  \end{center}
\end{figure}

We focus on showing the results of the case with $W_{0}=-10^{-5}$, $A=10$, and  $\kappa=2\pi$, for 
two representative initial field values $\phi_{\rm initial}$. Figure \ref{Fielddynamics} represents 
the evolution of the mean $\langle \phi \rangle$ and the variance $ \sqrt{ \langle\delta\phi^{2} \rangle }$ 
as a function of the scale factor for $\phi_{\text{initial}}=2.2$ (top) 
and $\phi_{\text{initial}}=2.2105$ (bottom), where $\phi_{\text{max}}=2.2105818...$. 
The solid black line indicates the field value at the inflection point of the potential. 

We do not see that the fluctuations grow for most of the initial values of the field 
($\phi_{\text{initial}}\lesssim 2.2$) as expected from the Floquet analysis, where the case for 
$\phi_{\text{initial}}= 2.2$ is shown in the top panels of Fig.~\ref{Fielddynamics}.

On the other hand, we see that the fluctuations grow by one order of magnitude at $a(t)\sim 6$ for 
$\phi_{\text{initial}}=2.2105$ shown in the bottom panels of Fig.~\ref{Fielddynamics}. This is because, 
when the field starts at very near the local maximum, it stays in the negative-curvature 
region of the potential for a sufficiently long time, so that the Hubble parameter becomes small enough to 
cause the aforementioned tachyonic instability. In addition, the fluctuations slightly experience 
another boost from $a(t)\sim 15$ due to parametric resonance. The growth of the fluctuations itself 
eventually comes to a halt because parametric resonance soon becomes inefficient when the amplitude of 
the fluctuations becomes almost as large as that of the homogenous mode, and the non-linear interactions 
by mode-mode couplings would be important thereafter. Then the amplitude starts to decrease 
due to the expansion of the Universe.

\begin{figure}[h]
  \begin{center}
    \begin{tabular}{cc}
      \begin{minipage}{0.45\hsize}
        \begin{center}
          \includegraphics[width=\textwidth]{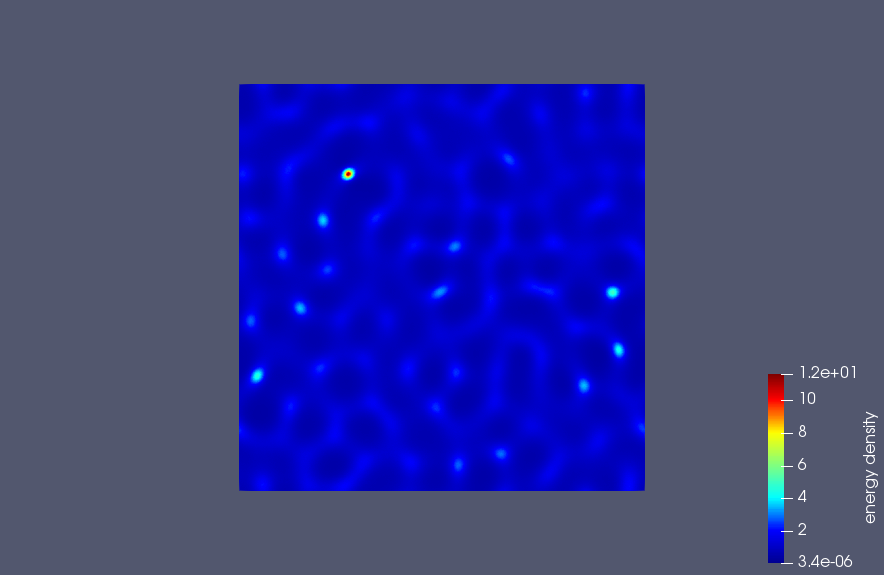} 
        \end{center}
      \end{minipage}
      &
      \begin{minipage}{0.45\hsize}
        \begin{center}
          \includegraphics[width=\textwidth]{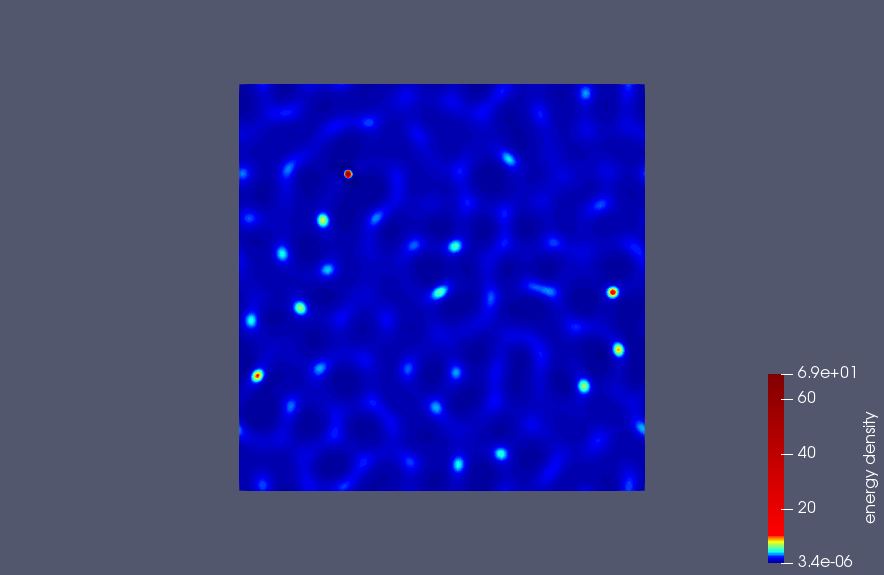}
        \end{center}
      \end{minipage}\\
       & \\
       \begin{minipage}{0.45\hsize}
        \begin{center}
          \includegraphics[width=\textwidth]{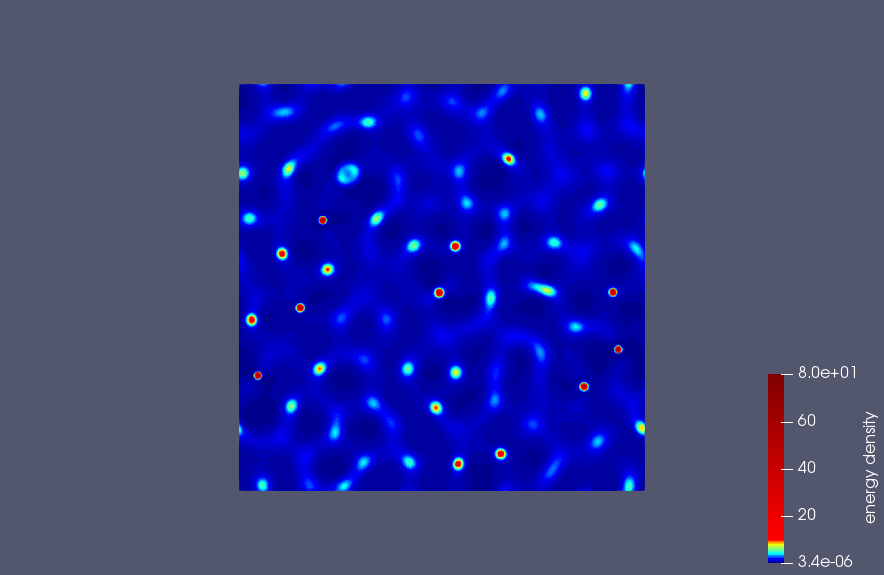}
        \end{center}
      \end{minipage}
      &
       \begin{minipage}{0.45\hsize}
        \begin{center}
          \includegraphics[width=\textwidth]{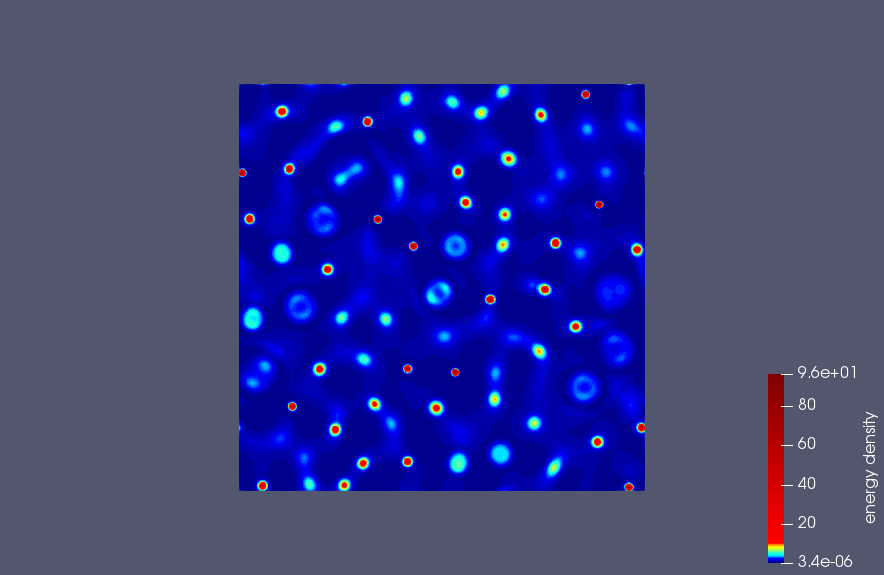}
        \end{center}
      \end{minipage}
    \end{tabular}
    \caption{Energy density distribution $\rho/\langle \rho \rangle$ in the two dimensional 
    lattice simulation for $W_{0}=-10^{-5}$, $A=10$, $\kappa=2\pi$ and $\phi_{\text{initial}}=2.2105$.
    The top left, top right, bottom left and bottom right panels correspond to the results 
    at $a=24.5$, $a=27.4$, $a=32.6$ and $a=44.2$, respectively. The red regions indicate
    $\rho/\langle \rho \rangle \geq 10$.}
    \label{W5simulation}
  \end{center}
\end{figure}

Next, we show in Fig.~\ref{W5simulation} the spatial distribution of the energy density 
of the K\"{a}hler modulus,
\begin{equation}
    \rho=\frac{1}{2}\dot{\phi}^{2}+\frac{1}{2a^{2}}|\nabla \phi|^{2}+V(\phi),
\end{equation}
normalized by the average density $\langle \rho \rangle$ at $a=24.5$ (top left), $a=27.4$ (top right), 
$a=32.6$ (bottom left) and $a=44.2$ (bottom right) for $\phi_{\text{initial}}=2.2105$. 
As can be seen, localized regions of overdensity are produced. They dominate the system well after 
the growth of the fluctuations has stopped and seems to be so even after the end of the simulation 
$a\sim70$. Since they could be a long-lived stable configuration, we can regard them as oscillons.
We confirm the formation of oscillons for 
\begin{equation}
    2.2103 \lesssim \phi_{\text{initial}} < \phi_{\text{max}},
\end{equation}
which is a very narrow region near the local maximum of the potential. 

We also perform numerical simulations for other values of the model parameters $W_{0}, A$, and $\kappa$ 
within the range of \eqref{parameterrange}, and obtain the same results that the oscillons can only form 
for a very narrow region of $\phi_{\text{initial}}$ near the local maximum of the potential. 
In addition, we check those cases having uplifting terms with different powers of $\sigma$, 
namely $D/\sigma, D/\sigma^{2}$, and $D/\sigma^{3}$, and find the same results with small 
quantitative but not qualitative differences. Therefore, we can safely say that severe fine-tuning 
for $\phi_{\text{initial}}$ is required for the oscillon formation in the KKLT model with the parameter 
range of \eqref{parameterrange}.

\section{Conclusion and discussion}
\label{sec:Conclusion-and-discussion}


We have revisited to consider the possibility of the oscillon formation in KKLT model following 
the previous work of \cite{Antusch:2017flz}. In there, they have concluded that the growth of 
the fluctuations is caused mostly by parametric resonance and that oscillons can be generated for 
$\phi_{\text{initial}}$ relatively close to the potential minimum. However, they set the initial
value of the Hubble parameter as $H_{\text{initial}}=\sqrt{V(\phi_{\text{initial}})/3}$, which seems
to be unnaturally small when the field begins to move towards the minimum of the potential.

In this paper, we have instead adopted the initial conditions as 
$H_{\text{initial}}=m_{\text{eff}}(\phi_{\rm initial})$, where $m_{\rm eff}$ is defined in (\ref{meff}), 
since it is this condition that the field starts rolling down the potential. This leads to the 
suppression of the growth of fluctuations, and the formation of oscillons is rather difficult.

By reconsidering the parameters thoroughly, we have thus come to a rather different conclusion:
a large portion of the growth of the fluctuations, even if it does exist, is caused by tachyonic instability, 
and oscillons can only form if the volume modulus is initially placed at very near the local maximum, which requires severe fine-tuning.

\section*{Acknowledgements}
This work was supported by JSPS KAKENHI Grant Nos. 17H01131 (M.K.) and 17K05434 (M.K.), MEXT KAKENHI Grant Nos. 15H05889 (M.K.), World Premier International Research Center Initiative (WPI Initiative), MEXT, Japan, and JSPS Research Fellowships for Young Scientists Grant No. 19J12936 (E.S.).
%

\bibliography{KKLToscillon}
\end{document}